\documentclass[journal]{IEEEtran}
\usepackage[pagewise, switch]{lineno}
\usepackage[hidelinks]{hyperref}
\usepackage{orcidlink}
\usepackage{cite}
\usepackage{tabularx}
\usepackage{float}
\usepackage{graphicx}
\usepackage{caption}
\usepackage{subcaption}
\usepackage{pgfplots}
\pgfplotsset{compat=newest}
\usepackage{url}
\usetikzlibrary{plotmarks}
\usetikzlibrary{arrows.meta}
\usepgfplotslibrary{patchplots}
\usepackage{grffile}
\usepackage{amsmath}
\usepackage{amssymb}

\ifCLASSINFOpdf
\else
\fi
\begin{document}
%\title{Systematic Review on Evaluation Methods for Gesture Generation in Embodied Conversational Agents}
\title{A Review of Evaluation Practices of Gesture Generation in Embodied Conversational Agents}


\author{Pieter Wolfert \orcidlink{0000-0002-7420-7181},
        Nicole Robinson \orcidlink{0000-0002-7144-3082}
        and~Tony~Belpaeme \orcidlink{0000-0001-5207-7745}% <-this % stops a space
\thanks{P. Wolfert and T. Belpaeme are with IDLab,
Ghent University - imec, Technologiepark 126, 9052, Gent Belgium, e-mail: pieter.wolfert@ugent.be. This research received funding from the Flemish Government (AI Research Program) and the Flemish Research Foundation grant no. 1S95020N.}% <-this % stops a space
\thanks{N. Robinson is with Monash University, Department of Electrical and Computer Systems Engineering, Turner Institute for Brain and Mental Health, Faculty of Engineering, Faculty of Medicine, Nursing and Health Sciences, Victoria, Australia

}% <-this % stops a space
\thanks{Manuscript received ...; revised ...}}

\markboth{\copyright 2022 IEEE}%
{Shell \MakeLowercase{\textit{et al.}}: Bare Demo of IEEEtran.cls for IEEE Journals}

\IEEEpubid{10.1109/THMS.2022.3149173 \copyright~2022 IEEE}



\maketitle

\begin{abstract}
% Embodied Conversational Agents (ECA) are known in many forms, they come in virtual avatars and physical agents such as humanoid robots. 
% They are often equipped with nonverbal behavior, since this improves human communication. 
% A specific form of nonverbal behavior, are co-speech gestures.
% There have been many studies on co-speech gestures for ECAs, yet these studies are not comparable, since a variety of objective and subjective evaluation methods are used, and there is no golden standard on which instruments to use. 
% In this paper we review 23 studies which are focused on the use of different co-speech gestures in ECAs, and we look at the evaluation methods used in these studies. 
% These studies are extracted from two databases (IEEE Xplore and Web of Science), with Google Scholar as an additional source. 
% We focus purely on papers that include an ECA with human hands, multiple gestures, a social human-agent scenario, user study and a direct or indirect rating by humans. 
% Additionally, we review the literature on beat gesture generation systems, where there is only one type of gesticulation generated. 
% In total, 1170 participants have taken part in studies reviewed in this paper. 
% Our aim is to provide an overview on the evaluation tools used in this field. 
% We show that the field requires stricter tools for the evaluation of co-speech gesture systems, and provide directions for improvement. 
Embodied conversational agents (ECA) are often designed to produce nonverbal behavior to complement or enhance their verbal communication. One such form of nonverbal behavior is co-speech gesturing, which involves movements that the agent makes with its arms and hands that are paired with verbal communication. Co-speech gestures for ECAs can be created using different generation methods, divided into rule-based and data-driven processes, with the latter gaining traction because of the increasing interest from the applied machine learning community. However, reports on gesture generation methods use a variety of evaluation measures, which hinders comparison. To address this, we present a systematic review on co-speech gesture generation methods for iconic, metaphoric, deictic, and beat gestures, including reported evaluation methods. We review 22 studies that have an ECA with a human-like upper body that uses co-speech gesturing in social human-agent interaction. This includes studies that use human participants to evaluate performance. We found most studies use a within-subject design and rely on a form of subjective evaluation, but without a systematic approach. We argue that the field requires more rigorous and uniform tools for co-speech gesture evaluation, and formulate recommendations for empirical evaluation, including standardized phrases and example scenarios to help systematically test generative models across studies. Furthermore, we also propose a checklist that can be used to report relevant information for the evaluation of generative models, as well as to evaluate co-speech gesture use. 

\end{abstract}

% Note that keywords are not normally used for peerreview papers.https://www.overleaf.com/project/5f06fa9beb9a740001142e03
\begin{IEEEkeywords}
human-robot interaction, virtual interaction, human-computer interface, social robotics
\end{IEEEkeywords}


\IEEEdisplaynontitleabstractindextext

% \fi
\IEEEpeerreviewmaketitle

\section{Introduction}

\IEEEPARstart{H}{uman} communication involves a large nonverbal component, with some suggesting that a large portion of communicative semantics is drawn from non-linguistic elements of face-to-face interaction \cite{knapp2013nonverbal}. Nonverbal behavior can be broken down into several elements, such as posture, gestures, facial expressions, gaze, proxemics, and haptics (i.e., touch during communicative interactions). All these elements convey different types of meaning, which can complement or alter the semantic component of communication. Even minimal elements can provide a marked contribution to the interaction. For example, eye blinking with head nodding has been found to influence the duration of a response in a Q\&A session between human subjects and a robot \cite{Homke2018}. 

A significant component involved in nonverbal communication is the use of gestures --movements of the hands, arms, or body-- to emphasize a message, communicate an idea, or express a sentiment \cite{knapp2013nonverbal}. Humans often use gestures in daily life, such as to point at objects in our visual space, or to signal the size of an object. Co-speech gestures are gestures that accompany speech. 
McNeill \cite{mcneill1992hand} categorized four kinds of co-speech gestures: iconic gestures, metaphorical gestures, beat gestures, and deictic gestures. Iconic and metaphorical gestures both carry meaning and are used to visually enrich our communication \cite{kendon1980gesticulation}. An iconic gesture can be an up and down movement to indicate, for example, the action of slicing a tomato. Instead, a metaphoric gesture can involve an empty palm hand that is used to symbolize `presenting a problem'. In other words, metaphoric gestures have an arbitrary relation to the concept they communicate, and iconic gestures have a form that is visually related to the concept being communicated. Iconic and metaphoric gestures not only differ in terms of content and presentation, but are also processed differently in the brain \cite{straube2011differentiation}. Beat gestures do not carry semantic meaning, and they are often used to emphasize the rhythm of speech. Beat gestures have been shown to both  facilitate speech and word recall \cite{lucero2014beat, igualada2017beat} and are the most frequent type of gesture \cite{mcneill1992hand, chui2005topicality, kong2015coding}. Finally, deictic gestures are used to point out elements of interest or to communicate directions. Not only do they enhance spoken communication, they also facilitate learning \cite{lucca2018communicating}. The remainder of this introduction covers gesture research in ECAs, evaluation methods, review aim, and objectives. 

\IEEEpubidadjcol

\begin{figure}
    \centering
    \includegraphics[width=0.42\textwidth]{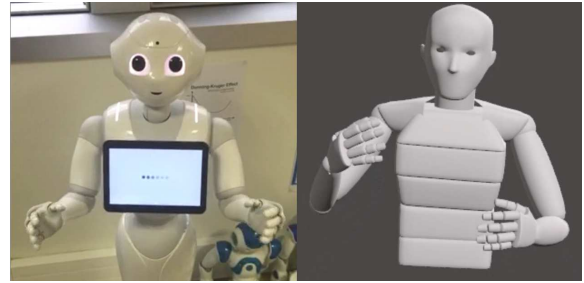}
    \caption{A Pepper robot (left) \cite{pandey2018mass} and a virtual avatar (right) \cite{alexanderson2020style} using their arms, hands, and torso to complement their speech with co-speech gestures.}
    \label{fig:pepperavatar}
\end{figure}

\subsection{Gesture Use in Human-Machine Interaction}
As nonverbal behavior plays an important role in human-human interaction, researchers put substantial efforts into the generation of nonverbal behavior for ECAs. ECAs, such as social robots today, can display a range of nonverbal behaviors, including the ability to make gesture-like movements \cite{bartneck2020human,breazeal2005effects,saunderson2019robots}. The use of co-speech gestures in communication with humans by ECAs can influence the perception and understanding of the conveyed message \cite{bremner2009conversational, allmendinger2010social}. 
For example, participants recalled more facts from a narrative told by an ECA, when the ECA made use of deictic and beat gestures compared to when the ECA did not make use of gesticulation \cite{huang2013modeling, huang2014learning}. As another example, humans are more willing to cooperate when an ECA showed appropriate gesturing (consisting of deictic, iconic, and metaphoric gestures) in comparison to when an ECA did not use gestures or when the gestures did not match the verbal utterances \cite{salem2013err}. Gestures are particularly salient in humanoid robotics, i.e., when the ECA is physically embodied. Robots can be perceived to be more persuasive when they combine gestures with other interactive social behaviors, such as eye gaze, in comparison with when they do not use either of these techniques \cite{ham2015combining,chidambaram2012designing,ghazali2018influence,ghazali2019assessing}. This demonstrates the impact nonverbal behavior from ECAs can have on people and its importance for consideration in human-agent interactions. 

Over the years, Artificial Intelligence (AI) powered systems have been used for the generation of communicative gestures. Gesture generation engines typically rely on matching language and gesture, given that the rhythm and semantic content signaled through gestures are highly correlated with the verbal utterance \cite{mcneill1992hand}. Early examples of ECA gesture generation relied on rule-based systems to generate gestures and nonverbal behavior, e.g., \cite{cassell1994}. For example, the BEAT system for generating nonverbal behavior can autonomously analyze input text on a linguistic and contextual level, and the system assigns nonverbal behaviors, such as beat and iconic gestures, based on predefined rules \cite{cassell2004beat}. A notable initiative was the Behavior Markup Language (BML), which provided a unified multimodal behavior generation framework \cite{kopp2006towards}. BML was used to describe physical behavior in an XML format and could be coupled with rule-based generation systems. To catch all aspects of nonverbal behavior generation, BML was aimed to not only integrate gesturing but also other forms such as body pose, head nodding, and gaze. 

Instead of relying on hand-coding, gesture generation systems can also be created from human conversational data, known as the data-driven approach \cite{levine2009real, bergmann2009gnetic}. These data-driven methods have predominantly relied on neural networks for synthesizing gestures. Paired with the rise of deep learning techniques, data-driven methods are capable of unprecedented generalization, an invaluable property when generating high dimensional temporal output. 
% Neural networks require large quantities of training data, and hold promise for producing high quality results. 
Data-driven approaches using neural networks are capable of generating more dynamic and unique gestures, but this does heavily depend on the available training data and the type of neural networks that are used. 
Some approaches learn a mapping from acoustic features of speech signals to gesture \cite{kucherenko2019analyzing, hasegawa2018evaluation}.
Audio signal-based methods are now much better at creating dynamic and fluent beat gestures, whereas text-based methods show an improved generation of iconic and metaphoric gestures. 
%These gesture generation methods have demonstrated the increase in availability of training data, capacity to create dynamic gestures and reduced the technical challenge with the use of current deep learning methods. 
However, relying on only acoustic features of the speech audio means that semantic details are lost, hence these approaches often only generate beat gestures. 
Recent work by Kucherenko et al. \cite{kucherenko2020gesticulator} combines neural networks for beat gesture generation with sequential neural networks for generating iconic gestures, dispensing with the need for a rule-based hybrid approach. Yoon et al. \cite{yoon2019robots}, trained an encoder-decoder neural network on combinations of subtitles and human poses extracted from public TED(x) videos. This allowed the network to learn a relationship between written language, extracted from the video's subtitles, and gesture and was used to generate beat and iconic gestures for a humanoid robot. 
However, an in-depth evaluation of the different categories of gestures generated by the system was not part of the study. This method was a notable advance in gesture generation, given that videos contain a wealth of human conversational data and are abundantly available. 
%On top of that, the network did not need to know specific social conventions to learn this relationship.
% To overcome this problem, researchers have tried a combination of learning and rule based systems, also known as hybrid systems.
The data used to build data-driven gesture generation can vary, where some use data collected from many individuals \cite{yoon2019robots}, others make use of data sets containing a single actor \cite{ferstl2019multi}. 
%%Subsection on Gesture Generation Methods

\subsection{Objective and Subjective Methods for Gesture Evaluation}
A central component for any method that can generate human-like behavior is the ability to evaluate the quality of the generated signals. To date, researchers make use of a variety of different methods to evaluate gesture generation systems. One way is to use objective evaluations, often consisting of metrics for the joint speed, joint trajectories, jerk, or the Frechet Gesture Distance \cite{yoon2020speech}. The objective metrics that are often reported are not necessarily the same metrics that are used to train neural networks. 
Loss functions only tell how close the generated stimuli are to the ground truth, and they do not provide information on whether the generated motion is dynamic or natural enough.
Others include subjective evaluations, which consist of a user study, where human participants evaluate the performance of the gestures used by the ECA.
Examples of dimensions on which the performance is evaluated, are, for example, the perceived naturalness of the generated motion, the perceived appropriateness of the gestures' timing, `speech-gesture correlation' or 'naturalness' \cite{ishi2018speech, levine2009real}. 
These are often evaluated using several items in one Likert Scale. In human-robot interaction \cite{yoon2019robots}, researchers have used questionnaires for general robot evaluation, such as the Godspeed questionnaire, or a selected subselection of items from such instruments. 
The Godspeed questionnaire can evaluate the perception of ECAs in a non-domain-specific measurement, and quantifies the human likeness, animacy, likability, and perceived intelligence of ECAs \cite{bartneck2008measuring}. 
Other methods measure the effect that the gesticulation of an ECA has on the user, such as listener's comprehension and recall of spoken material \cite{huang2013modeling, huang2014learning}.
In recent work by Ferstl et al. \cite{ferstl2021evaluating}, study designs and strategies for mitigating the impact of hand tracking loss in virtual reality are compared. In their experiments, they show the importance of asking the `right' question through comparing several evaluation strategies.
However, for the evaluation of generated co-speech gestures in ECAs, a standardized and validated evaluation methodology does not exist. 

As objective and subjective measures are central to assessing the quality of the generated communicative behavior, standardized evaluation methods and a uniform way of reporting measures will help to improve the quality of the field.
 
%The Godspeed questionnaire %:  \begin{quote}
%    ``proposes  a  series  of  questionnaires  to  measure  the  users’ perception of robots'' \cite{bartneck2008measuring}
%\end
\subsection{Review Aim and Objectives}
%The described collection of objective and subjective methods often mix direct and indirect measuring of factors important to gesture generation such as evaluation of the naturalness, semantic correctness and timing of co-speech gestures. 
%These items are often evaluated through the use of questionnaires, but where some ask in a direct form whether a gesture looks natural, others use an indirect way of questioning. However, they can contribute notable insight by indirect measurement of people's perception of the ECA, including robots. 
Given the importance that gestures can have on human-machine interaction, the ability to effectively identify and evaluate the appropriateness of gestures is vital.  
%Instead, the Godspeed questionnaire is intended for usage in HRI, with Likert items for anthropomorphism, animacy, likeability and perceived intelligence. 
However, there is no standardized generation and evaluation protocol available for the field of co-speech gesture generation for ECAs. A standardized questionnaire, measure, or protocol would make comparing work drawn from different sources more effective and would allow for more reliable reporting of results to demonstrate improvement over time. 
%To create a standardized method of gesture generation, the first step is knowing how to design systematic reporting methods. 
The completion of a comprehensive review and analysis of previous work in the field will support in understanding what has been accomplished so far and help establish a proposed protocol with systematic reporting methods that can be used for more robust evaluation of gesture generation methods, and their resulting gestures. 

In this paper, we present a systematic review that followed the Preferred Reporting Items for Systematic Reviews and Meta-Analyses (PRISMA) protocol \cite{moher2010preferred} to identify and assess evaluation methods used in co-speech gestures. 
We consider this review timely given that work in co-speech gesture generation is expanding, new techniques are emerging for creating novel gesture sets, and no systematic evaluation method has been provided to date. Central to this review, we have three research questions.

\begin{enumerate}
    \item What methods are used to evaluate co-speech gesture generation?
    \item Which methods can be considered the most effective for assessing co-speech gestures?
    \item What methods and related metrics should be adapted to create a standardized evaluation or reporting protocol?
\end{enumerate}

% The first question is: "What evaluation methods are used for evaluating co-speech gesture generation?"
% The second question: "What evaluation methodology appears to be the most effective for creating naturally occurring co-speech gestures (i.e. objective or subjective)?"
% The last and third question: "How can this inform a recommended standardised evaluation protocol and a set of metrics to match to for performance?"

These research questions will be used to formulate advice on how to make use of objective and subjective metrics to evaluate co-speech gesture performance of ECAs, including creating a standardized testing and reporting method. 

\section{Methods}
\subsection{Search Strategy}
This review focuses on evaluation studies of co-speech gesture generation methods for embodied conversational agents. Three databases were consulted for data extraction: IEEE Explore, Web of Science, and Google Scholar. IEEE Explore was selected given that it captures a substantial number of publications in computer science and engineering. Web of Science and Google Scholar were used because they provide access to multiple databases with a wide coverage extending beyond computer science and engineering. Data and record extraction occurred on April 8, 2020, and on June 25, 2020, to collect new records. Two authors conducted independent data extraction steps to reduce the chance of relevant papers being missed from the review, which included inter-rater checks on the included records. The databases were queried using four different keyword combinations, where the search engine would add `AND' between keywords: 1) “gesture generation for social robots”, 2) “co speech gesture generation”, 3) “non verbal gesture generation”, and 4) “nonverbal behavior generation”. 

\subsection{Eligibility – Inclusion and Exclusion}
%include the criteria and study selection
The following inclusion criteria were used:
\begin{enumerate}
    \item The ECA paper must report on gesture generation on either a robot or an embodied agent.
    \item The ECA system must be humanoid in nature, with one or two human-like arms and/or hands that can be used to gesture information or messages to the human. 
    \item The ECA system must display multiple gestures (i.e., a minimum of 2 different gestures, one of which must be a beat, iconic, metaphoric or deictic gesture). 
    \item Gestures created by the ECA system must be those that would be seen during a multi-modal social interaction. %human-agent scenario. %agent used to be robot but that seems to narrow.
    \item The ECA paper must report on a user study (i.e., not evaluated using technical collaborators or authors) in a laboratory, in the wild, or performed remotely through online platforms.
    \item The ECA system must be evaluated by a human rater on its performance (either directly or indirectly).
\end{enumerate}

To narrow down our search results, we used the following exclusion criteria: 

\begin{enumerate}
    \item The paper contains a non-humanoid agent that lacks a typical human-like hand for making a gesture.
    \item The paper does not have a clear focus on evaluation of co-speech gestures, i.e., secondary measures that is less than 50\% of the paper. 
    \item The paper only covers beat gesture generation.
    \item The paper is either unpublished, a doctoral dissertation, a review, a technical paper or pre-print. 
    \item The paper is not written in English. 
\end{enumerate}

Extracted records that only included beat gesture generation were recorded but excluded from the main analysis, as these records rely on audio inputs for the generation of beat gestures. Hence, these beat gesture generation systems do not take semantic information into account. Instead, a separate analysis outside the PRISMA protocol is provided to consider work on beat gestures only, as we do consider the work on beat gesture generation important. 
%Unpublished and other survey work such as doctoral dissertations, reviews, technical papers, pre-prints were not included in the search protocol. Only papers published in English were considered for this review. Papers could have been published in any given year. Authors were not directly contacted for any unpublished works to be included in this review. %Backward and forward searches were conducted on the final papers selected for the main analysis. 

\section{Results}
In this section, we discuss the results of our literature search. First, we discuss the found articles, followed by a discussion on the usage of different ECAs. Then, we discuss the characteristics of participant samples in experiments, the design of the experiments, and the use of objective and subjective evaluations. At the end, we present the results of our analysis of papers that only incorporated beat gesture generation.  
%\begin{figure}
%    \centering
%    \begin{tikzpicture}
%\begin{axis}[
%    title={Articles by year},
%    ybar stacked,
%    xlabel={Year},
%    ylabel={Frequency},
%    ymin=0, ymax=5,
%    minor y tick num = 1,
%    area style,
%	symbolic x coords={
%	2007, 2008, 2009, 2010, 2011, 2012, 2013, 2014, 2015, 2016, 2017, 2018, 2019}
%	]
%	\addplot coordinates { (2007, 0) (2008, 1) (2009, 0) (2010, 1) (2011, 0) (2012, 0) (2013, 1) (2014, 1) (2017, 1) %(2018, 1) (2019, 0)};
%    \addplot coordinates { (2007, 1) (2008, 0) (2009, 0) (2010, 1) (2011, 1) (2012, 4) (2013, 4) (2014, 1) (2017, 0) %(2018, 2) (2019, 2)};
%\legend{avatars,robots}
%\end{axis}
%
%\end{tikzpicture}
%\caption{Histogram of publication per year in the final selection.}
%    \label{fig:histogram}
%\end{figure}

\subsection{Selected Articles}
The initial search conducted across three separate databases resulted in 295 papers% and a total of 52 papers were extracted. 243 papers were taken from other sources (mainly Google Scholar) 
, which contained 92 duplicate records. A total of 203 papers were screened for their titles and abstracts for an initial exclusion step, resulting in 113 papers being omitted for not meeting all the inclusion criteria. The 90 remaining papers were assessed in detail by reviewing the main text for eligibility. The 68 non-eligible papers met one or more exclusion criteria, and were therefore discarded. This resulted in 22 papers that met all inclusion criteria and none of the exclusion criteria. Figure 2 shows the PRISMA flow chart with the results of this process. Extracted information from the manuscripts included publication year, venue, design and conditions, method of generation, objective metrics, subjective metrics, type of ECA, evaluation type (online, in the wild, or in a laboratory), participants, characteristics of participants, and other important notes related to the experiment. 

\begin{figure}
    \centering
    \includegraphics[width=0.50\textwidth]{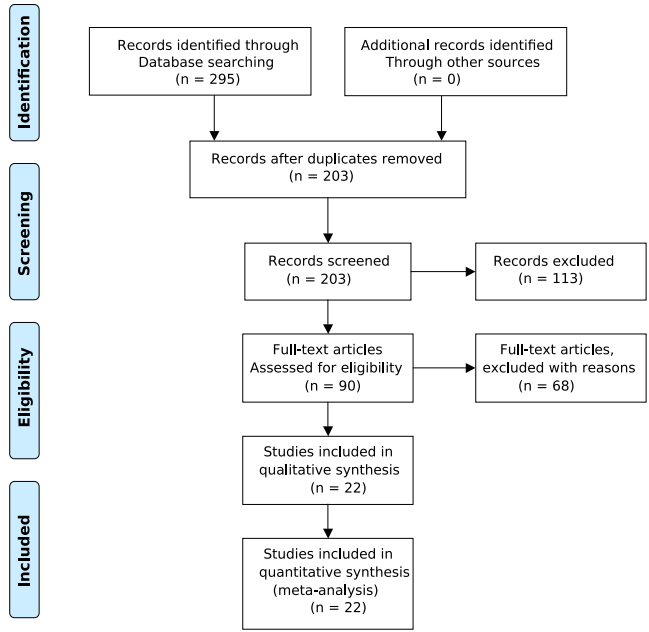}
    \caption{PRISMA Flow Chart}
    \label{fig:my_label}
\end{figure}

\subsection{Embodied Conversational Agents}
In the 22 included studies, 16 studies (73\%) used different human-like robots, such as NAO (n = 3, 14\%), ASIMO (n = 3, 14\%) or Wakamaru (n = 2, 9\%). Only 6 (27\%) reported the use of a virtual agent (\textit{viz.} \cite{ishii2018generating, neff2008gesture, xu2014compound, mlakar2013tts, levine2010gesture, rojc2017tts}). All the virtual agents were modelled in 3D as a virtual human, and there were no consistent features across the agents between studies. Of the 6 studies, 4 used female avatars \cite{rojc2017tts, mlakar2013tts,  xu2014compound, ishii2018generating}, 1 used a male avatar \cite{neff2008gesture} and 1 study used both\cite{levine2010gesture}. Half of the studies that used avatars, showed only the upper body \cite{rojc2017tts, mlakar2013tts, neff2008gesture}, whereas the other half showed full-body avatars \cite{levine2010gesture, xu2014compound, ishi2018speech}. Specific descriptions of the hands were not provided in all the studies that used avatars. In 19 (87\%) studies, the ECA performed iconic gestures, combined with other gestures \cite{yoon2019robots, salem2012generation, salem2013closing, ishii2018generating, ishi2018speech, aly2013model, mlakar2013tts, shimazu2018generation, perez2019part, le2012evaluating, ng2010synchronized, rojc2017tts, le2012common, salem2011friendly, bennewitz2007fritz, huang2013modeling, huang2014learning, salem2013err}. 

Metaphoric gestures, with other gestures, are used in 17 (77\%) studies \cite{yoon2019robots, salem2012generation, salem2013closing, ishii2018generating, ishi2018speech, neff2008gesture, xu2014compound, aly2013model, mlakar2013tts, le2012evaluating, ng2010synchronized, rojc2017tts, le2012common, salem2011friendly, huang2014learning, huang2013modeling, salem2013err}. 
Deictic gestures, with other gesture types, play a key role in 13 (59\%) of the reviewed studies \cite{yoon2019robots, salem2012generation, salem2013closing, ishii2018generating, ishi2018speech, mlakar2013tts, kim2012gesture, le2012evaluating, ng2010synchronized, rojc2017tts, le2012common, salem2011friendly, bennewitz2007fritz, huang2013modeling, huang2014learning, salem2013err}. Lastly, 17 (77\%) studies included iconic, metaphoric \textit{and} beat gestures \cite{yoon2019robots, ishii2018generating, ishi2018speech, neff2008gesture, xu2014compound, aly2013model, shimazu2018generation, perez2019part, kim2012gesture, le2012evaluating, levine2010gesture, ng2010synchronized, rojc2017tts, le2012common, bennewitz2007fritz, huang2013modeling, huang2014learning}. Half of the studies had the ECA perform `random gestures' that were included in the evaluation (i.e., gestures that had no alignment between gestures and speech). Other studies (n = 4) had the ECA present the user with a variety of different nonverbal behavior schemes, such as gestures that were based on text, speech, or a combination of the two\cite{shimazu2018generation, ishii2018generating, perez2019part, salem2013err}. 

\subsection{Participants}
The number of participants per study ranged from 13 to 250 in total (mean = 50, SD = 50, median = 35). In these papers, 19 (86\%) were conducted in the laboratory, and 3 (14\%) were conducted either online through Amazon Mechanical Turk (AMT) (n = 2) and 1 during an exhibition (i.e., `in the wild'). For the 12 (54\%) studies that did report the mean age of the participants, the mean reported age across all studies was 30.10 years of age (SD = 6.6). The remaining 11 (46\%) did not provide demographic data for gender and age. Relating to trial location, 16 (73\%) of studies were performed outside English-speaking countries, with the top 3 countries being Germany (n = 5), Japan (n = 3), and France (n = 3). For participant recruitment, 6 (27\%) of the studies reported the use of university students --a so-called \textit{convenience sample}-- to evaluate gesture generation. Table I provide a more detailed overview of the different studies, countries of origin, and characteristics. 

\begin{table*}[!ht]
\caption{Participants in Studies}
\centering
%\begin{tabular}{c|c|c|c|c|c|c}
\renewcommand{\arraystretch}{1.2}
\begin{tabular}{@{}lllllll@{}}
\hline
\textbf{Study} & \textbf{Country} & \textbf{Gender} & \textbf{Mean Age (SD)} & \textbf{$N$} & \textbf{Characteristics} & \textbf{Lab/Remote Evaluation} \\
\hline
\cite{yoon2019robots} & South Korea & 23M/23F & 37 (-) & 46 & 45 USA, 1 Australia & AMT \\
\cite{perez2019part} & Spain & - & - & 50 & Non-native English Speakers & In Lab \\
\cite{ishii2018generating} & Japan & - & - & 10 & Age + Gender not specified & In Lab \\
\cite{ishi2018speech} & Japan & - & - & 20 & Age + Gender not specified & In Lab \\
\cite{shimazu2018generation} & Japan & - & - & 13 & - & In Lab \\
\cite{rojc2017tts} & Slovenia & 22M/8F & - & 30 & - & In Lab \\
\cite{xu2014compound} & U.S.A. & - & - & 250 & One `worker' per comparison & AMT \\
\cite{huang2014learning} & U.S.A. & 16M/13F & 22.62 (4.35) & 29 & Convenience Sample & In Lab \\
\cite{salem2013closing} & Germany & 10M/10F & 28.5 (4.53) & 20 & Native German Speakers & In Lab \\
\cite{aly2013model} & France & 14M/7F & 21-30 & 21 & Convenience Sample & In Lab \\
\cite{mlakar2013tts} & Slovenia & 23M/7F & 26.73 (4.88) & 30 & Convenience Sample & In Lab \\
\cite{huang2013modeling} & U.S.A. & 16M/16F & 24.34 (8.64) & 32 & Convenience Sample & In Lab \\
\cite{salem2013err} & Germany & 30M/32F & 30.90 (9.82) & 62 & Convenience Sample & In Lab \\
\cite{salem2012generation} & Germany & 30M/30F & 31 (10.21) & 60 & Native German Speakers & In Lab \\
\cite{kim2012gesture} & South Korea & - & - & 65 &  & In Lab \\
\cite{le2012evaluating} & France & 36M/27F & 37 (12.14) & 63 & Convenience Sample & In Lab \\
\cite{le2012common} & France & - & - & 63 & French Speakers & In Lab \\
\cite{salem2011friendly} & Germany & 20M/20F - 20M/20F & 31.31 (10.55)/31.54(10.96) & 81 & Two Studies & In Lab \\
\cite{levine2010gesture} & U.S.A. & 21M/14F & 23 (-) & 35 & Convenience Sample & In Lab \\
\cite{ng2010synchronized} & U.S.A. & - & - & 54 & - & In Lab \\
\cite{neff2008gesture} & U.S.A. & 20M/6F & 24-26 (-) & 26 & Non-experts & In Lab \\
\cite{bennewitz2007fritz} & Germany & - & - & - & - & Exhibition\\
\hline
 
\end{tabular}
\end{table*}

\subsection{Research Experiment and Assessment}
In research design, 16 (68\%) of the studies used a within-subject design and 7 (32\%) used a between-subject design. 
Most (n = 18, 82\%) studies invited participants to a university research laboratory to have an interaction with an ECA.
Other methods used AMT (n = 2, 9\%). 
With use in 9 (41\%) studies, `naturalness' was the most common metric for evaluation in generated gestures. This was followed by synchronization (n = 6, 27\%), likability (n = 4, 18\%), and human-likeness (n = 2, 9\%). 2 studies (9\%) \cite{salem2013closing, xu2014compound} asked participants to choose which audio track matched best with a given generated gesture sequence. 9 (41\%) studies made use of models that learn to generate co-speech gestures. 
When assessing generated gestures, 16 (73\%) studies used questionnaires as a tool to evaluate ECA gesture performance. Only 1 study \cite{salem2013closing} included a previous iteration of their gesture model for evaluation.  4 studies (18\%) used a ground truth as part of the gesture generation evaluation. 3 studies (13\%) relied on pairwise comparisons, such as two or more videos put side by side with the user selecting the video that best matches with the speech audio, e.g., \cite{perez2019part, levine2010gesture, ng2010synchronized}. Other evaluation methods involved robot performance, e.g., \cite{huang2013modeling, huang2014learning}. 

\subsection{Objective and Subjective Evaluation}
Table II provides a summary of studies that involved objective evaluation. It also includes the type of agents that were used, as well as the number of speakers in a dataset (when applicable) and the setting of the speakers in the conversation. Only 5 studies (23\%) involved some form of objective evaluation metrics as a key method in their evaluation. Other metrics included variations on the mean squared error (MSE) (n = 1, 4.5\%) between the generated and ground truth gestures, and qualitative analyses of joint velocities and positions (n = 2, 9\%). In total, 10 (45\%) studies used a data-driven generation method, but only 3 studies (14\%) reported outcomes of their objective metrics used for tuning their models. Only 3 (14\%) studies reported the results of their objective metrics relating to their model performance. 7 studies (32\%) relied on data featuring single speakers. In addition to that, 7 studies (32\%) relied on data showing 2 or more speakers. The remainder did not report on the setting of the data or the number of speakers in their dataset. 

Table III provides a detailed overview of study design, conditions, and subjective evaluation methods. Fewer studies used between-group design (n = 6, 27\%) compared to within-group design (n = 16, 73\%). Most were evaluated using questionnaires (n = 16, 73\%) followed by pairwise comparisons (n = 3, 14\%) and other methods (n = 4, 18\%) such as preference matching (matching audio with video) and recalling facts from a story told by the agent. 

\begin{table*}[!ht]
\centering
\caption{Objective Evaluation Methods}
%\begin{tabular}{c|c|c|c|c|c}
\begin{tabular}{@{}llllll@{}}
\hline
     \textbf{Study} & \textbf{Generation Method} & \textbf{Objective Metrics} & \textbf{Agent} & \textbf{\#N Speakers} & \textbf{Setting}  \\
     \hline
\cite{yoon2019robots} & Data Driven & Variation on Mean Squared Error & NAO & 1295 & Single \\
\cite{perez2019part} & Rule Based & - & REEM-C & 2 & Single\\
\cite{ishii2018generating} & Data Driven & - & Virtual Agent (3D) & 24 & Conversation (two)\\
\cite{ishi2018speech} & Data Driven & - & Android Erica & 8 & Conversation (three)\\
\cite{shimazu2018generation} & Data Driven & Log-likelihood of generated motion & Pepper & 119 & Single \\
\cite{rojc2017tts} & Hybrid & - & Virtual Agent (3D)  & 5 & Multiple\\
\cite{xu2014compound} & Rule Based & - & Virtual Agent (3D) & 5 & Single\\
\cite{huang2014learning} & Data Driven & - & Wakamaru & 16 & Conversation (two)\\
\cite{salem2013closing} & Rule Based & Qualitative Analysis of Joint Positions & ASIMO & - & - \\
\cite{aly2013model} & Rule Based & - & NAO & - & -  \\
\cite{mlakar2013tts} & Data Driven & - & Virtual Agent (3D) & 4 & Multiple\\
\cite{huang2013modeling} & Rule Based & - & Wakamaru & 8 & Conversation (two)\\
\cite{salem2013err} & Rule Based & - & ASIMO  & - & - \\
\cite{salem2012generation} & Rule Based & Qualitative Analysis of Joint Positions & ASIMO & - & -  \\
\cite{kim2012gesture} & Rule Based & - & Industrial Service Robot & 1 & Single \\
\cite{le2012evaluating} & Rule Based & - & NAO & - & -  \\
\cite{le2012common} & Rule Based & - & NAO & - & -  \\
\cite{salem2011friendly} & Rule Based & - & ASIMO & - & -  \\
\cite{levine2010gesture} & Data Driven & Cost Function on Kinematic Parameters & Virtual Agent (3D) & 1 & Conversation (two) \\
\cite{ng2010synchronized} & Rule Based & - & ASIMO & 4 & Single\\
\cite{neff2008gesture} & Data Driven & - & Virtual Agent (3D) & 2 & Single \\
\cite{bennewitz2007fritz} & Rule Based & - & Fritz & - & - \\
\hline 
\end{tabular}
\end{table*}

%\subsection{Subjective Evaluations}
\begin{table*}[!ht]
\centering
\caption{Subjective Evaluation Methods}
\label{table3}
\begin{tabularx}{\textwidth}{@{}llXXXX@{}}
\hline
\textbf{Study} & \textbf{Design} & \textbf{Conditions} & \textbf{Gesture Types} & \textbf{Evaluation} & \textbf{Questionnaire items} \\
\hline
\cite{yoon2019robots} & Within-subject & Ground truth, proposed method, nearest neighbors, random or manual & Iconic, Beat, Deictic, Metaphoric & Questionnaire & Anthropomorphism, Likability, Speech-gesture correlation \\
\cite{perez2019part} & Within-subject & Part-of-Speech-Based, Prosody-Based, Combined & Iconic, Beat & Pairwise + Questionnaire & Timing, Appropriateness, Naturalness \\
\cite{ishii2018generating} & Within-subject & None, Random, Proposed Method & Iconic, Beat, Deictic, Metaphoric & Questionnaire & Naturalness of Movement, Consistency in utterance and movement, likability, humanness \\
\cite{ishi2018speech} & Within-subject & No hand motion, Direct Human mapping, Text-based gestures, Text-based + prosody-based gestures & Iconic, Beat, Deictic, Metaphoric & Questionnaire & Human-likeness, Gesture-speech suitability, Gesture-Naturalness, Gesture-Frequency, Gesture-timing \\
\cite{shimazu2018generation} & Within-subject & Ground truth, seq2seq, seq2seq(model) + semantic, seq2seq\_tts + semantic & Iconic, Beat & Questionnaire & Naturalness, Skill of presentation, Utilization of gesture, Vividness, Enthusiasm \\
\cite{rojc2017tts} & Within-subject & Text+Speech (no avatar), Gestures & Iconic, Beat, Deictic, Metaphoric & Questionnaire & Content Match, Synchronization, Fluidity, Dynamics, Density, Understanding, Vividness \\
\cite{xu2014compound} & Within-subject & Hands never go into relax position, hands always go into rest position & Beat, Metaphoric & Match preference & N.A. \\
\cite{huang2014learning} & Between-subject & Learning-based, unimodal, random, conventional & Iconic, Beat, Deictic, Metaphoric & Questionnaire + Retelling Performance & Immediacy, Naturalness, Effectiveness, Likability, Credibility \\
\cite{salem2013closing} & Within-subject & Old version, new version of model & Iconic, Deictic, Metaphoric & Match preference & N.A. \\
\cite{aly2013model} & Within-subject & Introverted versus Extraverted Robot, Adapted Speech and Behavior versus Adapted Speech & Iconic, Beat, Metaphoric & Questionnaire & 24 questions on personality, interaction with the robot, speech, and gesture synchronization and matching \\
\cite{mlakar2013tts} & Between-subject & Virtual avatar versus iCub robot & Iconic, Deictic, Metaphoric & Questionnaire & Content Matching, Synchronization, Fluidness, Speech-Gesture Matching, Execution Speed, Amount of Gesticulation \\
\cite{huang2013modeling} & Between-subject & Number of gestures, randomly selected & Iconic, Beat, Deictic, Metaphoric & Questionnaire + Retelling Performance & Naturalness, Competence, Effective use of Gestures \\
\cite{salem2013err} & Between-subject & Unimodal (speech only), congruent multimodal, incongruent multimodal & Iconic, Deictic, Metaphoric & Questionnaire & Human likeness, Likability, Shared Reality, Future Contact Intentions \\
\cite{salem2012generation} & Between-subject & Unimodal versus multimodal (speech + gestures) in a kitchen task & Iconic, Deictic, Metaphoric & Questionnaire & Gesture Quantity, Gesture Speed, Gesture Fluidity, Speech-Gesture Content, Speech-Gesture Timing, Naturalness \\
\cite{kim2012gesture} & Within-subject & - & Deictic, Beat & Questionnaire & Suitability of Gestures, Synchronization, Scheduling \\
\cite{le2012evaluating} & Within-subject & Synchronized Gestures, not Synchronized Gestures, Gestures with Expressivity, Gestures without Expressivity & Iconic, Beat, Deictic, Metaphoric & Questionnaire & Synchronization, Naturalness, Expressiveness, Contradictiveness, Gestures are complementary, Gesture-speech Redundancy \\
\cite{le2012common} & Within-subject & One Condition & Iconic, Beat, Deictic, Metaphoric & Questionnaire & Speech-Gesture Synchronization, Expressiveness, Naturalness \\
\cite{salem2011friendly} & Between-subject & Study 1: Unimodal versus Multimodal; Study 2: Same & Iconic, Deictic, Metaphoric & Questionnaire & Appearance, Naturalness, Liveliness, Friendliness \\
\cite{levine2010gesture} & Within-subject & Generated versus Ground Truth & Iconic, Beat & Pairwise & - \\
\cite{ng2010synchronized} & Within-subject & 4 studies: Audio vs Wrong Audio; Excited vs Calm Gestures; Low Expressivity, Medium Expressivity, High Expressivity; Slow Gesticulation, Medium Gesticulation, Fast Gesticulation & Iconic, Beat, Deictic, Metaphoric & Pairwise & - \\
\cite{neff2008gesture} & Within-subject & Speaker 1, speaker 2 & Beat, Metaphoric & Match style to speaker & - \\
\cite{bennewitz2007fritz} & - & - & Iconic, Beat, Deictic & Public Exhibition & -\\
\hline
\end{tabularx}
\end{table*}
 
\subsection{Additional Results – Beat Gestures}
Research work that focused on \textit{only} beat gesture generation was excluded from the main analysis. Methods used to evaluate the performance of beat gesture generation systems in ECAs were similar to those used in work on semantic gesture generation. 10 papers were selected that met the criteria \cite{kucherenko2019analyzing, ondras2020audio, kim2012automated, bremner2009beat, chiu2014gesture, fernandez2014gesture, levine2009real, wolfert2019should, takeuchi2017speech, kipp2007towards}. A total of 7 (70\%) studies mentioned the number of participants, with a total of 236 participants. Only 4 (40\%) mentioned statistics on age and gender. Of the 10 studies, 4 (40\%) were performed in a lab, and 5 (50\%) online or via AMT. 1 study was evaluated in an exhibition. As beat gesture generation mostly relied on prosody information, 8 (80\%) studies used a data-driven approach. Only 4 of the 8 studies that relied on data-driven methods reported their metrics used for an objective evaluation, with either the average position error (APE) or the MSE. 7 (70\%) of papers ran their evaluation on a virtual avatar or stick figure with no discernible face. The subjective evaluations performed in these studies were similar to studies that included more gesture categories. 
6 (60\%) used a post-experiment questionnaire to assess the quality of the generated gestures by the ECA. 30\% relied on pairwise comparisons and 1 (10\%) relied on the time spent with focused attention on an ECA \cite{bremner2009beat}.
All studies (n = 10) relied on a within-subject evaluation. The questionnaire items that were used the most: `naturalness' (n = 4, 40\%) and `time consistency' (n = 4, 40\%). 

\section{Principal Findings and Implications}
In this section, we examine the above observations in more detail and discuss implications for gesture generation methods. Due to the high variation and diversity in the experiments presented in the main analysis, a meta-analysis of the experiments' results will not be provided.  

\subsection{Participant Sample}
More than half of the studies involved in the main analysis did not report details on the raters, such as the average age, gender, or cultural background. This is a challenge for knowing the generalizability of the findings to larger samples, or its appropriateness for a particular cultural and geographical context. Many studies (30\%) used participants that were readily available, for example from a higher education campus. However, such a \textit{convenience sample} of students is not representative of the general population and may result in a sample of a predominant young adult cohort from higher socioeconomic backgrounds, which might bias the results \cite{peterson2014convenience}. Subsequently, the evaluation of gestures generated from models represents a more narrow cultural and social viewpoint, and some gestures that are acceptable and natural in other cultures may have been misrepresented or rated poorly in the evaluation process from the use of a more restricted sample. 
%The reported participant sample also came from diverse geographical locations, which are known to have unique gesture styles and communication methods. This would have influenced the accuracy, naturalness and perceived performance ratings of gesture generation \cite{lafrance1978cultural}. While it is a strength to have diversity, these studies cannot be easily pooled together due to variation in each trial, which creates a cultural challenge to know if highly rated gestures represent a specific viewpoint. This causes difficulty in drawing more general conclusions on gesture generation and evaluation practices conducted to date.

\subsection{Recruitment and Trial Location}
The use of online workers, through services such as AMT or Prolific, does have its merits. Large amounts of data can be collected for a modest budget and in a very short period of time, and it can reach participants from different global regions with very diverse backgrounds. In addition, studies have shown that crowd-sourced data can be of comparable quality to lab-based studies \cite{buhrmester2011amazon}. Given that the majority of users on AMT are US-based, it is important that studies report the cultural background and country of residence for their participants \cite{moss2020ethical}. Although a recent study showed there might be no difference between studies for the evaluation of gesture generation in ECAs in the lab and on AMT, it is important to include attention checks and response quality control mechanisms, and to report on these \cite{jonell2020can}. 

\subsection{Experimental Set-up and Assessment}
In the main analysis, 14 (65\%) studies relied on a within-subject design, which helps to evaluate iterations of gestures over multiple exposures, introduces less variation in the participant scores, and requires fewer participants to achieve sufficient statistical power. 
It is, however, somewhat problematic that not all studies relied on ground truth comparisons. A ground truth condition typically is a recording of gestures by a human with corresponding speech audio, which are then compared to computer-generated gestures. Human ground truth can serve as a concrete baseline, and this should score the highest on scales for appropriateness and naturalness, providing a clear comparison with other evaluation scores. Several studies also involved random movement generation as a control condition. Random movement is interpreted in different ways, some take random samples from their data set, which are then put on top of original speech \cite{yoon2019robots}, or insert random parameters for generating gestures \cite{huang2014learning}. Random gestures are an important control condition for this type of work, ensuring that people are not simply attributing meaning to every gesture seen in the experiment, whether it was a relevant co-speech gesture or not. Overall, we note that the quality of the experimental set-up for gesture generation and evaluation was moderate. 

\subsection{Evaluation Methods}
The reviewed literature did not show a consistent use of evaluation metrics for gestures, with different research groups focusing on features of interest to them specifically. In most cases, evaluation methods such as questionnaires were used for assessing the quality of co-speech gestures in ECAs \cite{yoon2019robots, ishii2018generating, shimazu2018generation, le2012common}. Different questionnaires did extract information around similar outcomes, but there was no gold standard for questionnaires, or agreement on a single questionnaire to evaluate the perception of generated gestures. Many items were conflated in a single dimension, which causes an evaluation to miss detail. Questionnaires often involved the use of Likert scales, which sometimes are incorrectly used \cite{schrum2020four}, such as failing to report internal consistency, except for \cite{huang2013modeling, huang2014learning}. Objective evaluations were also highly varied, from using MSE to reporting on histograms with joint velocities and positions. 
%Overall, evaluation methods are still preliminary and maturing, and the presented studies did not allow for a recommended instrument or methodology to be nominated. 

\section{Recommendations for Gesture Evaluation}
In the previous section, we discussed the principal findings of our literature review on evaluation used in co-speech gesture generation. Following our findings and our experience, we provide recommendations for researchers working in this field. First, we give more general recommendations, coupled with examples from other, relevant fields. 
Secondly, we propose an additional method of evaluation, for which we provide sentences and scenarios. Lastly, we introduce a checklist that researchers can incorporate in their future work, to improve the level of reporting on datasets, methodology, and results.

\subsection{Participant Sample}
As mentioned in the previous section, many studies fail to report on the details of the participant samples. Additionally, not all participant samples reflect the data on which models or systems are trained. We recommend subjective evaluations with participants from diverse populations and backgrounds, reflecting the data on which models or systems are trained. 

Some work is more focussed on equipping virtual agents with gesticulation, whereas others take it a step further and use their methodology to drive nonverbal behavior in social robots. Often, intermediate evaluation is overlooked, which can potentially lead to unwanted results when these engines are used in an interactive scenario. We recommend that participant evaluation is conducted -when feasible- before putting the model in production or when using the model on a new data-set, ensuring better validity and relevance when deployed for human social interaction. 

\subsection{Experimental setup}
In this section, we cover recommendations relating to the conditions, design of studies, and measurements.

\begin{figure}
    \centering
\definecolor{ao(english)}{HTML}{648FFF}%
\definecolor{applegreen}{HTML}{DC267F}%
\definecolor{newgreen}{HTML}{FFB000}%
\begin{tikzpicture}

\begin{axis}[%
width=2.18in,
height=2.18in,
at={(1.728in,1.449in)},
scale only axis,
xmin=30,
xmax=100,
xlabel style={font=\color{white!15!black}},
xlabel={Human-likeness},
ymin=30,
ymax=100,
ylabel style={font=\color{white!15!black}},
ylabel={Appropriateness},
title={Subjective Ratings GENEA Challenge 2020},
axis background/.style={fill=white},
axis x line*=bottom,
axis y line*=left,
legend style={at={(0.725,0.113)}, anchor=south west, legend cell align=left, align=left, fill=none, draw=none, line width=3pt}
]
\addplot [fill=ao(english), color=ao(english)]
  table[row sep=crcr]{%
70	79\\
75	79\\
75	83\\
70	83\\
70	79\\
};
\addlegendentry{Ground Truth}

\addplot [fill=applegreen, color=applegreen]
  table[row sep=crcr]{%
70	53\\
75	53\\
75	59\\
70	59\\
70	53\\
};
\addlegendentry{Mismatched}

\addplot [color=newgreen, fill=newgreen]
  table[row sep=crcr]{%
44	38\\
49	38\\
49	41\\
44	41\\
44	38\\
};
\addlegendentry{Data-driven models}

\addplot [color=newgreen, fill=newgreen, forget plot]
  table[row sep=crcr]{%
53	35\\
58	35\\
58	40\\
53	40\\
53	35\\
};
\addplot [color=newgreen, fill=newgreen, forget plot]
  table[row sep=crcr]{%
35	31\\
41	31\\
41	37\\
35	37\\
35	31\\
};
\addplot [color=newgreen, fill=newgreen, forget plot]
  table[row sep=crcr]{%
50	40\\
55	40\\
55	45\\
50	45\\
50	40\\
};
\addplot [color=newgreen, fill=newgreen, forget plot]
  table[row sep=crcr]{%
55	48\\
60	48\\
60	52\\
55	52\\
55	48\\
};
\addplot [color=newgreen, fill=newgreen, forget plot]
  table[row sep=crcr]{%
57	46\\
61	46\\
61	50\\
57	50\\
57	46\\
};
\addplot [color=newgreen, fill=newgreen, forget plot]
  table[row sep=crcr]{%
47	44\\
51	44\\
51	49\\
47	49\\
47	44\\
};
\addplot [color=black, dotted, forget plot]
  table[row sep=crcr]{%
0	0\\
100	100\\
};
\end{axis}
\end{tikzpicture}%
    \caption{Human-likeness and appropriateness subjective measurements comparisons between data-driven models and the ground truth from the GENEA 2020 Challenge. Adapted from \cite{kucherenko2021large}.}
    \label{fig:geneachallenge}
\end{figure}
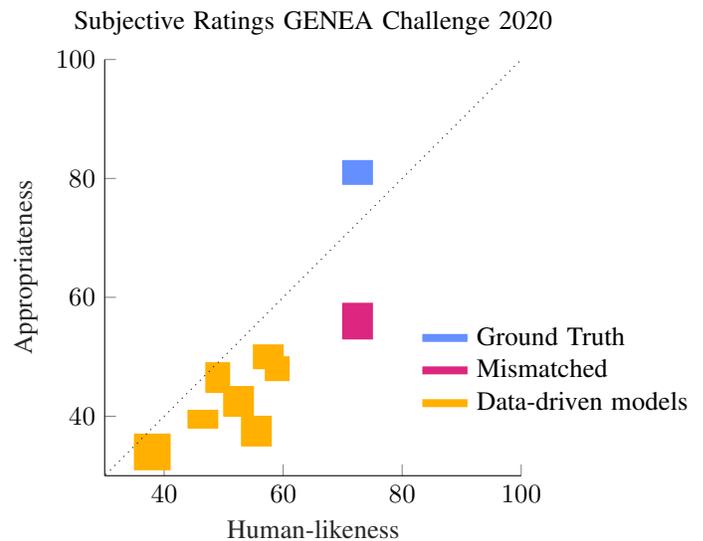

The cornerstone of each subjective evaluation is to compare the output of a system to the ground truth. This ground truth condition must contain both motion and audio. Another condition that can shed light on a system’s performance, is a random or mismatched condition, in which real motion is put on top of a different audio track. An interesting example of this is the subjective evaluation that was part of the GENEA 2020 Challenge, part of the International Conference on Intelligent Virtual Agents (IVA), and to our knowledge, the first of its kind in this field \cite{kucherenko2021large}. In this challenge, multiple data-driven co-speech generators were compared to two baseline systems. A crowd-sourced subjective evaluation was part of this challenge, for which the results on `appropriateness’ and `human-likeness’ are displayed in Figure 3. Here, we see that ground truth is scored higher than the submitted systems on both dimensions and can function as a proper baseline. As for human-likeness, the mismatched condition offers an intriguing result: it does still look as human-like as the ground truth, yet it is scored much lower on appropriateness. Both a ground truth condition and a mismatched condition can function as a sanity check when being compared to the output of a system. 

Most studies analyzed for this review ask participants to rate individual stimuli. This can be substantiated with more rigor using the contrastive approach, also known as A/B testing or side-by-side testing \cite{kohavi2017online}. With such an approach, two or more stimuli are presented at the same moment, and a user is asked to either rate both stimuli or to select the preferred stimulus. In a recent study by the authors, these two types of a contrastive approach were tested, as we wanted to find out whether one of the two contrastive approaches should be preferred \cite{wolfert2021rate}. In one condition, participants were asked to make a choice between two videos (pairwise comparison) or to rate both videos. The authors found that when evaluating many conditions, an approach that makes use of rating scales is to be preferred over using pairwise comparisons. However, pairwise comparisons are a lot faster and less cognitively demanding on participants \cite{weijters2010effect}. 

Many studies evaluate the performance of their approach in a one-way fashion: videos are put online and participants are asked to evaluate individual videos. However, the need for proper gesticulation in ECAs is often tied to how humans communicate with each other. We recommend (when feasible) evaluating these systems in an interactive scenario, given that it is often the aim of researchers to eventually use ECAs in interactive scenarios. This might require additional engineering, such as creating systems that can also deal with synthetic speech (and thus with entirely new input), and creating dialogues to be used in an interactive scenario. However, by using an interactive scenario to evaluate an ECAs performance, it becomes possible to record and annotate interactions for indirect measurements, which we will discuss in the next paragraph. 

A common way of evaluating stimuli is to ask for ratings on certain dimensions on a 5 or 7 point scale. Table III shows us the richness in terms of questionnaire items used for subjective evaluations. These items can also be seen as `direct’ items since they are used for direct measurement on a certain dimension. Frequently used items for this are ‘naturalness’, ‘human-likeness’, ‘appropriateness’, or ‘likability’. Our recommendation here, when one wants to rely on direct measurements only, is that subjective evaluations cover specific dimensions: naturalness, human-likeness, fluency, appropriateness, or intelligibility. Ideally, these dimensions are scored on a 5 or 7 point scale (as these tend to provide more reliable results than larger scales \cite{dawes2008data}). In addition to direct measurements, we would like to make the case for using a more indirect way of measuring. Examples of indirect measurements are the time it takes to complete a task (task completion), recall rate (recall of facts when letting an ECA tell a story), eye contact and gaze, or response duration (in a question-answering session). For example, task completion is an often-used proxy to estimate effectiveness in human-computer interaction \cite{jordan2020introduction}, and might serve a similar role in our domain. The recall rate has already been used to evaluate gestures \cite{huang2013modeling, huang2014learning}, but could play a more important role in future interactive evaluations. Eye contact, gaze, or response duration are good proxies to estimate a user's engagement, and taking engagement into account has worked well for other domains \cite{nakano2010estimating, lemaignan2016real}. The level of engagement could in turn be a good predictor of how effective an ECA's gesticulation is. However, the drawback of using indirect ways of measuring, is that some of these approaches require annotating video recordings of experimental sessions with multiple raters. 

\subsection{Qualitative Analysis of Model Output}
Data-driven models are often trained on a combination of speech-audio and text. Whereas some systems rely on one speaker (as is the case with systems submitted for the GENEA 2020 Challenge), others rely on multiple speakers. When data-driven systems are capable of generating gestures independent of a specific input voice, it becomes possible to use synthetic text-to-speech as input. This in turn makes it possible to present new data and to qualitatively analyze the performance of models on this new data. We propose a new task that takes entirely new sentences (and text-to-speech output when necessary) as input for gesture generation models. The output then needs to be analyzed for the occurrence of gesture categories. For example, for the sentence “I was throwing a ball”, a model might generate an iconic gesture for the word ‘ball’. We have crowdsourced a set of sentences and scenarios that can be used for this task \footnote{\url{https://github.com/pieterwolfert/gesturegeneration-checklist}}. We propose that researchers take a subset of these as input and that they annotate the model's output for the occurrence of gesture categories. This approach can provide an insight into the richness and diversity of the output of these models. However, this task only works for systems that can work with either only input text or a combination of input text and synthetic speech audio. 

\subsection{Preferred reporting items for Gesture Generation Researchers}
To supplement the recommendations made in the previous sections, we offer a non-exhaustive list with preferred reporting items. These draw upon our observations of reporting and our research experiences (\cite{wolfert2019should, wolfert2021rate, kucherenko2021large}). Considering the items in the proposed list, researchers could further enhance the quality of their reporting. Our proposed list with items that would be worth including in future work is summarized in Table IV. It contains items we deem important to report in a scientific publication when working on gesture generation for both physical and non-physical agents. We hope that the use of this list will make it easier in the future to allow for more systematic evaluation and benchmarking.

\begin{table}[!ht]
\renewcommand{\arraystretch}{1.2}
\caption{Preferred reporting items for co-speech gesture evaluation}
\label{table4}
\begin{tabularx}{0.50\textwidth}{X}
\textbf{Embodied Conversational Agent}:\\
$\square$ ECA: Avatar/robot \\
%$\square$ Gesture generation model\\
$\square$ DOF (shoulder, elbow, wrist, hand, neck) \\
$\square$ Level of articulation of hands \\
\textbf{Demographics}: \\
$\square$ Recruitment method \\
$\square$ Sample size \\
$\square$ Age \\
$\square$ Gender distribution \\
$\square$ Geographical distribution \\
$\square$ Prior exposure with ECAs \\
$\square$ Language(s) spoken \\
\textbf{Gesture Generation Model}: \\
$\square$ Included generated gestures: [iconic, metaphorical, beat, deictic] \\
$\square$ Gesture generation model: [rule based, data driven, both, other] \\
$\square$ Gesture generation model link/repository\\
$\square$ (If not included – why not?) \\
\textbf{Gesture Generation Evaluation}: \\
$\square$ Context / application \\
$\square$ Evaluation method/questionnaire set \\
$\square$ Gestures annotated by human raters? [Yes/No] \\
$\square$ How many human raters were used?\\
$\square$ Inter-rater agreement \\
\textbf{Metrics}: \\
$\square$ Objective metrics [average jerk, distance between velocity histograms] \\
$\square$ Subjective metrics [human likeness, gesture appropriateness, quality, other] \\
\textbf{Training dataset}: \\
$\square$ Domain of dataset \\
$\square$ Length/size of dataset \\
$\square$ Gesture types annotated in the dataset \\
$\square$ Details on the actors in the dataset ($N$, language, conversation topic) \\
\textbf{Statistical analysis scripts}: \\
$\square$ Link to scripts
\end{tabularx}
\end{table}

\section{Conclusion}
We reviewed 22 studies on the generation and evaluation of co-speech gestures for ECAs, with a specific focus on evaluation methods. Three questions guided our review, with the first asking what methods are used to evaluate co-speech gesture generation. We found a large diversity of different methods, both objective and subjective, that were applied to the evaluation of generated co-speech gestures. Our main analysis found that many studies did not mention basic statistics on participant characteristics, few studies reported detailed evaluation methods, and there were no systematic reporting methods used for gesture generation and evaluation steps. Our second question asked which methodology is most effective for assessing co-speech gestures. From our review, we cannot conclude that one way of evaluating is to be preferred over another, and recommend making use of both objective and subjective methods. Our third and final question asked what methods and metrics should be adapted to create a standardized evaluation or reporting protocol. Our findings indicate that the field of gesture generation and evaluation would benefit from more experimental rigor and a shared methodology for conducting systematic evaluations, see e.g., \cite{hoffman2020primer, liu2021speech}. We offer questionnaire dimensions, a list with preferred items for designing and reporting studies, and new evaluation tasks, and call on the community to work towards a standardized protocol and questionnaire for the evaluation of systems that produce co-speech gestures. We hope that this work can contribute to further development of the field and that it will contribute to further advancements in terms of co-speech gesture generation in ECAs. 

\ifCLASSOPTIONcaptionsoff
  \newpage
\fi
\bibliographystyle{IEEEtran} 
\bibliography{Transactions-Bibliography/references.bib}

\begin{thebibliography}{10}
\providecommand{\url}[1]{#1}
\csname url@samestyle\endcsname
\providecommand{\newblock}{\relax}
\providecommand{\bibinfo}[2]{#2}
\providecommand{\BIBentrySTDinterwordspacing}{\spaceskip=0pt\relax}
\providecommand{\BIBentryALTinterwordstretchfactor}{4}
\providecommand{\BIBentryALTinterwordspacing}{\spaceskip=\fontdimen2\font plus
\BIBentryALTinterwordstretchfactor\fontdimen3\font minus
  \fontdimen4\font\relax}
\providecommand{\BIBforeignlanguage}[2]{{%
\expandafter\ifx\csname l@#1\endcsname\relax
\typeout{** WARNING: IEEEtran.bst: No hyphenation pattern has been}%
\typeout{** loaded for the language `#1'. Using the pattern for}%
\typeout{** the default language instead.}%
\else
\language=\csname l@#1\endcsname
\fi
#2}}
\providecommand{\BIBdecl}{\relax}
\BIBdecl

\bibitem{knapp2013nonverbal}
M.~L. Knapp, J.~A. Hall, and T.~G. Horgan, \emph{Nonverbal communication in
  human interaction}.\hskip 1em plus 0.5em minus 0.4em\relax Cengage Learning,
  2013.

\bibitem{Homke2018}
P.~H{\"{o}}mke, J.~Holler, and S.~C. Levinson, ``{Eye blinks are perceived as
  communicative signals in human face-to-face interaction},'' \emph{PLoS ONE},
  2018.

\bibitem{mcneill1992hand}
D.~McNeill, \emph{Hand and mind: What gestures reveal about thought}.\hskip 1em
  plus 0.5em minus 0.4em\relax University of Chicago press, 1992.

\bibitem{kendon1980gesticulation}
A.~Kendon, ``Gesticulation and speech: Two aspects of the,'' \emph{The
  relationship of verbal and nonverbal communication}, no.~25, p. 207, 1980.

\bibitem{straube2011differentiation}
B.~Straube, A.~Green, B.~Bromberger, and T.~Kircher, ``The differentiation of
  iconic and metaphoric gestures: Common and unique integration processes,''
  \emph{Human brain mapping}, vol.~32, no.~4, pp. 520--533, 2011.

\bibitem{lucero2014beat}
C.~Lucero, H.~Zaharchuk, and D.~Casasanto, ``Beat gestures facilitate speech
  production,'' in \emph{Proceedings of the Annual Meeting of the Cognitive
  Science Society}, vol.~36, no.~36, 2014.

\bibitem{igualada2017beat}
A.~Igualada, N.~Esteve-Gibert, and P.~Prieto, ``Beat gestures improve word
  recall in 3-to 5-year-old children,'' \emph{Journal of Experimental Child
  Psychology}, vol. 156, pp. 99--112, 2017.

\bibitem{chui2005topicality}
K.~Chui, ``Topicality and gesture in chinese conversational discourse,''
  \emph{LANGUAGE AND LINGUISTICS-TAIPEI-}, vol.~6, no.~4, p. 635, 2005.

\bibitem{kong2015coding}
A.~P.-H. Kong, S.-P. Law, C.~C.-Y. Kwan, C.~Lai, and V.~Lam, ``A coding system
  with independent annotations of gesture forms and functions during verbal
  communication: Development of a database of speech and gesture (dosage),''
  \emph{Journal of nonverbal behavior}, vol.~39, no.~1, pp. 93--111, 2015.

\bibitem{lucca2018communicating}
K.~Lucca and M.~P. Wilbourn, ``Communicating to learn: Infants’ pointing
  gestures result in optimal learning,'' \emph{Child development}, vol.~89,
  no.~3, pp. 941--960, 2018.

\bibitem{pandey2018mass}
A.~K. Pandey and R.~Gelin, ``A mass-produced sociable humanoid robot: Pepper:
  The first machine of its kind,'' \emph{IEEE Robotics \& Automation Magazine},
  vol.~25, no.~3, pp. 40--48, 2018.

\bibitem{alexanderson2020style}
S.~Alexanderson, G.~E. Henter, T.~Kucherenko, and J.~Beskow,
  ``Style-controllable speech-driven gesture synthesis using normalising
  flows,'' in \emph{Computer Graphics Forum}, vol.~39, no.~2.\hskip 1em plus
  0.5em minus 0.4em\relax Wiley Online Library, 2020, pp. 487--496.

\bibitem{bartneck2020human}
C.~Bartneck, T.~Belpaeme, F.~Eyssel, T.~Kanda, M.~Keijsers, and
  S.~{\v{S}}abanovi{\'c}, \emph{Human-robot interaction: An
  introduction}.\hskip 1em plus 0.5em minus 0.4em\relax Cambridge University
  Press, 2020.

\bibitem{breazeal2005effects}
C.~Breazeal, C.~D. Kidd, A.~L. Thomaz, G.~Hoffman, and M.~Berlin, ``Effects of
  nonverbal communication on efficiency and robustness in human-robot
  teamwork,'' in \emph{2005 IEEE/RSJ international conference on intelligent
  robots and systems}.\hskip 1em plus 0.5em minus 0.4em\relax IEEE, 2005, pp.
  708--713.

\bibitem{saunderson2019robots}
S.~Saunderson and G.~Nejat, ``How robots influence humans: A survey of
  nonverbal communication in social human--robot interaction,''
  \emph{International Journal of Social Robotics}, vol.~11, pp. 575--608, 2019.

\bibitem{bremner2009conversational}
P.~Bremner, A.~Pipe, C.~Melhuish, M.~Fraser, and S.~Subramanian,
  ``Conversational gestures in human-robot interaction,'' in \emph{2009 IEEE
  international conference on systems, man and cybernetics}.\hskip 1em plus
  0.5em minus 0.4em\relax IEEE, 2009, pp. 1645--1649.

\bibitem{allmendinger2010social}
K.~Allmendinger, ``Social presence in synchronous virtual learning situations:
  The role of nonverbal signals displayed by avatars,'' \emph{Educational
  Psychology Review}, vol.~22, no.~1, pp. 41--56, 2010.

\bibitem{huang2013modeling}
C.-M. Huang and B.~Mutlu, ``Modeling and evaluating narrative gestures for
  humanlike robots,'' in \emph{Robotics: Science and Systems}, 2013, pp.
  57--64.

\bibitem{huang2014learning}
------, ``Learning-based modeling of multimodal behaviors for humanlike
  robots,'' in \emph{2014 9th ACM/IEEE International Conference on Human-Robot
  Interaction (HRI)}.\hskip 1em plus 0.5em minus 0.4em\relax IEEE, 2014, pp.
  57--64.

\bibitem{salem2013err}
M.~Salem, F.~Eyssel, K.~Rohlfing, S.~Kopp, and F.~Joublin, ``To err is human
  (-like): Effects of robot gesture on perceived anthropomorphism and
  likability,'' \emph{International Journal of Social Robotics}, vol.~5, no.~3,
  pp. 313--323, 2013.

\bibitem{ham2015combining}
J.~Ham, R.~H. Cuijpers, and J.-J. Cabibihan, ``Combining robotic persuasive
  strategies: The persuasive power of a storytelling robot that uses gazing and
  gestures,'' \emph{International Journal of Social Robotics}, vol.~7, no.~4,
  pp. 479--487, 2015.

\bibitem{chidambaram2012designing}
V.~Chidambaram, Y.-H. Chiang, and B.~Mutlu, ``Designing persuasive robots: how
  robots might persuade people using vocal and nonverbal cues,'' in
  \emph{Proceedings of the seventh annual ACM/IEEE international conference on
  Human-Robot Interaction}, 2012, pp. 293--300.

\bibitem{ghazali2018influence}
A.~S. Ghazali, J.~Ham, E.~Barakova, and P.~Markopoulos, ``The influence of
  social cues in persuasive social robots on psychological reactance and
  compliance,'' \emph{Computers in Human Behavior}, vol.~87, pp. 58--65, 2018.

\bibitem{ghazali2019assessing}
------, ``Assessing the effect of persuasive robots interactive social cues on
  users’ psychological reactance, liking, trusting beliefs and compliance,''
  \emph{Advanced Robotics}, vol.~33, no. 7-8, pp. 325--337, 2019.

\bibitem{cassell1994}
J.~Cassell, C.~Pelachaud, N.~Badler, M.~Steedman, B.~Achorn, T.~Becket,
  B.~Douville, S.~Prevost, and M.~Stone, ``Animated conversation: Rule-based
  generation of facial expression, gesture \& spoken intonation for multiple
  conversational agents,'' in \emph{Proceedings of the 21st Annual Conference
  on Computer Graphics and Interactive Techniques}, ser. SIGGRAPH ’94.\hskip
  1em plus 0.5em minus 0.4em\relax New York, NY, USA: Association for Computing
  Machinery, 1994, p. 413–420.

\bibitem{cassell2004beat}
J.~Cassell, H.~H. Vilhj{\'a}lmsson, and T.~Bickmore, ``Beat: the behavior
  expression animation toolkit,'' in \emph{Life-Like Characters}.\hskip 1em
  plus 0.5em minus 0.4em\relax Springer, 2004, pp. 163--185.

\bibitem{kopp2006towards}
S.~Kopp, B.~Krenn, S.~Marsella, A.~N. Marshall, C.~Pelachaud, H.~Pirker, K.~R.
  Th{\'o}risson, and H.~Vilhj{\'a}lmsson, ``Towards a common framework for
  multimodal generation: The behavior markup language,'' in \emph{International
  workshop on intelligent virtual agents}.\hskip 1em plus 0.5em minus
  0.4em\relax Springer, 2006, pp. 205--217.

\bibitem{levine2009real}
S.~Levine, C.~Theobalt, and V.~Koltun, ``Real-time prosody-driven synthesis of
  body language,'' in \emph{ACM SIGGRAPH Asia 2009 papers}, 2009, pp. 1--10.

\bibitem{bergmann2009gnetic}
K.~Bergmann and S.~Kopp, ``Gnetic--using bayesian decision networks for iconic
  gesture generation,'' in \emph{International Workshop on Intelligent Virtual
  Agents}.\hskip 1em plus 0.5em minus 0.4em\relax Springer, 2009, pp. 76--89.

\bibitem{kucherenko2019analyzing}
T.~Kucherenko, D.~Hasegawa, G.~E. Henter, N.~Kaneko, and H.~Kjellstr{\"o}m,
  ``Analyzing input and output representations for speech-driven gesture
  generation,'' in \emph{Proceedings of the 19th ACM International Conference
  on Intelligent Virtual Agents}, 2019, pp. 97--104.

\bibitem{hasegawa2018evaluation}
D.~Hasegawa, N.~Kaneko, S.~Shirakawa, H.~Sakuta, and K.~Sumi, ``Evaluation of
  speech-to-gesture generation using bi-directional lstm network,'' in
  \emph{Proceedings of the 18th International Conference on Intelligent Virtual
  Agents}, 2018, pp. 79--86.

\bibitem{kucherenko2020gesticulator}
T.~Kucherenko, P.~Jonell, S.~van Waveren, G.~E. Henter, S.~Alexanderson,
  I.~Leite, and H.~Kjellstr{\"o}m, ``Gesticulator: A framework for
  semantically-aware speech-driven gesture generation,'' in \emph{Proceedings
  of the ACM International Conference on Multimodal Interaction}, 2020.

\bibitem{yoon2019robots}
Y.~Yoon, W.-R. Ko, M.~Jang, J.~Lee, J.~Kim, and G.~Lee, ``Robots learn social
  skills: End-to-end learning of co-speech gesture generation for humanoid
  robots,'' in \emph{2019 International Conference on Robotics and Automation
  (ICRA)}.\hskip 1em plus 0.5em minus 0.4em\relax IEEE, 2019, pp. 4303--4309.

\bibitem{ferstl2019multi}
Y.~Ferstl, M.~Neff, and R.~McDonnell, ``Multi-objective adversarial gesture
  generation,'' in \emph{Motion, Interaction and Games}, 2019, pp. 1--10.

\bibitem{yoon2020speech}
Y.~Yoon, B.~Cha, J.-H. Lee, M.~Jang, J.~Lee, J.~Kim, and G.~Lee, ``Speech
  gesture generation from the trimodal context of text, audio, and speaker
  identity,'' \emph{ACM Transactions on Graphics (TOG)}, vol.~39, no.~6, pp.
  1--16, 2020.

\bibitem{ishi2018speech}
C.~T. Ishi, D.~Machiyashiki, R.~Mikata, and H.~Ishiguro, ``A speech-driven hand
  gesture generation method and evaluation in android robots,'' \emph{IEEE
  Robotics and Automation Letters}, vol.~3, no.~4, pp. 3757--3764, 2018.

\bibitem{bartneck2008measuring}
C.~Bartneck, D.~Kulic, and E.~Croft, ``\BIBforeignlanguage{English}{Measuring
  the anthropomorphism, animacy, likeability, perceived intelligence, and
  perceived safety of robots},'' in
  \emph{\BIBforeignlanguage{English}{Proceedings of the 3rd ACM/IEEE
  international conference on Human robot interaction}}, 2008, annual ACM/IEEE
  International Conference on Human-Robot Interaction (HRI) 2008, HRI 2008 ;
  Conference date: 12-03-2008 Through 15-03-2008.

\bibitem{ferstl2021evaluating}
Y.~Ferstl, R.~McDonnell, and M.~Neff, ``Evaluating study design and strategies
  for mitigating the impact of hand tracking loss,'' in \emph{ACM Symposium on
  Applied Perception 2021}, 2021, pp. 1--12.

\bibitem{moher2010preferred}
D.~Moher, A.~Liberati, J.~Tetzlaff, D.~G. Altman \emph{et~al.}, ``Preferred
  reporting items for systematic reviews and meta-analyses: the prisma
  statement,'' \emph{Int J Surg}, vol.~8, no.~5, pp. 336--341, 2010.

\bibitem{ishii2018generating}
R.~Ishii, T.~Katayama, R.~Higashinaka, and J.~Tomita, ``Generating body motions
  using spoken language in dialogue,'' in \emph{Proceedings of the 18th
  International Conference on Intelligent Virtual Agents}, 2018, pp. 87--92.

\bibitem{neff2008gesture}
M.~Neff, M.~Kipp, I.~Albrecht, and H.-P. Seidel, ``Gesture modeling and
  animation based on a probabilistic re-creation of speaker style,'' \emph{ACM
  Transactions on Graphics (TOG)}, vol.~27, no.~1, pp. 1--24, 2008.

\bibitem{xu2014compound}
Y.~Xu, C.~Pelachaud, and S.~Marsella, ``Compound gesture generation: a model
  based on ideational units,'' in \emph{International Conference on Intelligent
  Virtual Agents}.\hskip 1em plus 0.5em minus 0.4em\relax Springer, 2014, pp.
  477--491.

\bibitem{mlakar2013tts}
I.~Mlakar, Z.~Ka{\v{c}}i{\v{c}}, and M.~Rojc, ``Tts-driven synthetic
  behaviour-generation model for artificial bodies,'' \emph{International
  Journal of Advanced Robotic Systems}, vol.~10, no.~10, p. 344, 2013.

\bibitem{levine2010gesture}
S.~Levine, P.~Kr{\"a}henb{\"u}hl, S.~Thrun, and V.~Koltun, ``Gesture
  controllers,'' in \emph{ACM SIGGRAPH 2010 papers}, 2010, pp. 1--11.

\bibitem{rojc2017tts}
M.~Rojc, I.~Mlakar, and Z.~Ka{\v{c}}i{\v{c}}, ``The tts-driven affective
  embodied conversational agent eva, based on a novel conversational-behavior
  generation algorithm,'' \emph{Engineering Applications of Artificial
  Intelligence}, vol.~57, pp. 80--104, 2017.

\bibitem{salem2012generation}
M.~Salem, S.~Kopp, I.~Wachsmuth, K.~Rohlfing, and F.~Joublin, ``Generation and
  evaluation of communicative robot gesture,'' \emph{International Journal of
  Social Robotics}, vol.~4, no.~2, pp. 201--217, 2012.

\bibitem{salem2013closing}
M.~Salem, S.~Kopp, and F.~Joublin, ``Closing the loop: Towards tightly
  synchronized robot gesture and speech,'' in \emph{International Conference on
  Social Robotics}.\hskip 1em plus 0.5em minus 0.4em\relax Springer, 2013, pp.
  381--391.

\bibitem{aly2013model}
A.~Aly and A.~Tapus, ``A model for synthesizing a combined verbal and nonverbal
  behavior based on personality traits in human-robot interaction,'' in
  \emph{2013 8th ACM/IEEE International Conference on Human-Robot Interaction
  (HRI)}.\hskip 1em plus 0.5em minus 0.4em\relax IEEE, 2013, pp. 325--332.

\bibitem{shimazu2018generation}
A.~Shimazu, C.~Hieida, T.~Nagai, T.~Nakamura, Y.~Takeda, T.~Hara, O.~Nakagawa,
  and T.~Maeda, ``Generation of gestures during presentation for humanoid
  robots,'' in \emph{2018 27th IEEE International Symposium on Robot and Human
  Interactive Communication (RO-MAN)}.\hskip 1em plus 0.5em minus 0.4em\relax
  IEEE, 2018, pp. 961--968.

\bibitem{perez2019part}
L.~P{\'e}rez-Mayos, M.~Farr{\'u}s, and J.~Adell, ``Part-of-speech and
  prosody-based approaches for robot speech and gesture synchronization,''
  \emph{Journal of Intelligent \& Robotic Systems}, pp. 1--11, 2019.

\bibitem{le2012evaluating}
Q.~A. Le and C.~Pelachaud, ``Evaluating an expressive gesture model for a
  humanoid robot: Experimental results,'' in \emph{Submitted to 8th ACM/IEEE
  International Conference on Human-Robot Interaction}, 2012.

\bibitem{ng2010synchronized}
V.~Ng-Thow-Hing, P.~Luo, and S.~Okita, ``Synchronized gesture and speech
  production for humanoid robots,'' in \emph{2010 IEEE/RSJ International
  Conference on Intelligent Robots and Systems}.\hskip 1em plus 0.5em minus
  0.4em\relax IEEE, 2010, pp. 4617--4624.

\bibitem{le2012common}
Q.~Le, J.~Huang, and C.~Pelachaud, ``A common gesture and speech production
  framework for virtual and physical agents,'' in \emph{ACM international
  conference on multimodal interaction}, 2012.

\bibitem{salem2011friendly}
M.~Salem, K.~Rohlfing, S.~Kopp, and F.~Joublin, ``A friendly gesture:
  Investigating the effect of multimodal robot behavior in human-robot
  interaction,'' in \emph{2011 Ro-Man}.\hskip 1em plus 0.5em minus 0.4em\relax
  IEEE, 2011, pp. 247--252.

\bibitem{bennewitz2007fritz}
M.~Bennewitz, F.~Faber, D.~Joho, and S.~Behnke, ``Fritz-a humanoid
  communication robot,'' in \emph{RO-MAN 2007-The 16th IEEE International
  Symposium on Robot and Human Interactive Communication}.\hskip 1em plus 0.5em
  minus 0.4em\relax IEEE, 2007, pp. 1072--1077.

\bibitem{kim2012gesture}
H.-H. Kim, Y.-S. Ha, Z.~Bien, and K.-H. Park, ``Gesture encoding and
  reproduction for human-robot interaction in text-to-gesture systems,''
  \emph{Industrial Robot: An International Journal}, 2012.

\bibitem{ondras2020audio}
J.~Ondras, O.~Celiktutan, P.~Bremner, and H.~Gunes, ``Audio-driven robot
  upper-body motion synthesis,'' \emph{IEEE Transactions on Cybernetics}, 2020.

\bibitem{kim2012automated}
J.~Kim, W.~H. Kim, W.~H. Lee, J.-H. Seo, M.~J. Chung, and D.-S. Kwon,
  ``Automated robot speech gesture generation system based on dialog sentence
  punctuation mark extraction,'' in \emph{2012 IEEE/SICE International
  Symposium on System Integration (SII)}.\hskip 1em plus 0.5em minus
  0.4em\relax IEEE, 2012, pp. 645--647.

\bibitem{bremner2009beat}
P.~Bremner, A.~G. Pipe, M.~Fraser, S.~Subramanian, and C.~Melhuish, ``Beat
  gesture generation rules for human-robot interaction,'' in \emph{RO-MAN
  2009-The 18th IEEE International Symposium on Robot and Human Interactive
  Communication}.\hskip 1em plus 0.5em minus 0.4em\relax IEEE, 2009, pp.
  1029--1034.

\bibitem{chiu2014gesture}
C.-C. Chiu and S.~Marsella, ``Gesture generation with low-dimensional
  embeddings,'' in \emph{Proceedings of the 2014 international conference on
  Autonomous agents and multi-agent systems}, 2014, pp. 781--788.

\bibitem{fernandez2014gesture}
A.~Fern{\'a}ndez-Baena, R.~Monta{\~n}o, M.~Antonijoan, A.~Roversi, D.~Miralles,
  and F.~Al{\'\i}as, ``Gesture synthesis adapted to speech emphasis,''
  \emph{Speech communication}, vol.~57, pp. 331--350, 2014.

\bibitem{wolfert2019should}
P.~Wolfert, T.~Kucherenko, H.~Kjelstr{\"o}m, and T.~Belpaeme, ``Should beat
  gestures be learned or designed? a benchmarking user study,'' in
  \emph{ICDL-EPIROB 2019 Workshop on Naturalistic Non-Verbal and Affective
  Human-Robot Interactions}, 2019, pp. 1--4.

\bibitem{takeuchi2017speech}
K.~Takeuchi, D.~Hasegawa, S.~Shirakawa, N.~Kaneko, H.~Sakuta, and K.~Sumi,
  ``Speech-to-gesture generation: A challenge in deep learning approach with
  bi-directional lstm,'' in \emph{Proceedings of the 5th International
  Conference on Human Agent Interaction}, 2017, pp. 365--369.

\bibitem{kipp2007towards}
M.~Kipp, M.~Neff, K.~H. Kipp, and I.~Albrecht, ``Towards natural gesture
  synthesis: Evaluating gesture units in a data-driven approach to gesture
  synthesis,'' in \emph{International Workshop on Intelligent Virtual
  Agents}.\hskip 1em plus 0.5em minus 0.4em\relax Springer, 2007, pp. 15--28.

\bibitem{peterson2014convenience}
R.~A. Peterson and D.~R. Merunka, ``Convenience samples of college students and
  research reproducibility,'' \emph{Journal of Business Research}, vol.~67,
  no.~5, pp. 1035--1041, 2014.

\bibitem{buhrmester2011amazon}
M.~Buhrmester, T.~Kwang, and S.~D. Gosling, ``Amazon's mechanical turk: A new
  source of inexpensive, yet high-quality, data?'' \emph{Perspectives on
  Psychological Science}, vol.~6, no.~1, pp. 3--5, 2011.

\bibitem{moss2020ethical}
A.~J. Moss, C.~Rosenzweig, J.~Robinson, and L.~Litman, ``Is it ethical to use
  mechanical turk for behavioral research? relevant data from a representative
  survey of mturk participants and wages,'' 2020.

\bibitem{jonell2020can}
P.~Jonell, T.~Kucherenko, I.~Torre, and J.~Beskow, ``Can we trust online
  crowdworkers? comparing online and offline participants in a preference test
  of virtual agents,'' in \emph{Proceedings of the 20th ACM International
  Conference on Intelligent Virtual Agents}, 2020, pp. 1--8.

\bibitem{schrum2020four}
M.~L. Schrum, M.~Johnson, M.~Ghuy, and M.~C. Gombolay, ``Four years in review:
  Statistical practices of {Likert} scales in human-robot interaction
  studies,'' in \emph{Companion of the 2020 ACM/IEEE International Conference
  on Human-Robot Interaction}, 2020, pp. 43--52.

\bibitem{kucherenko2021large}
T.~Kucherenko, P.~Jonell, Y.~Yoon, P.~Wolfert, and G.~E. Henter, ``A large,
  crowdsourced evaluation of gesture generation systems on common data: The
  genea challenge 2020,'' in \emph{26th International Conference on Intelligent
  User Interfaces}, 2021, pp. 11--21.

\bibitem{kohavi2017online}
R.~Kohavi and R.~Longbotham, ``Online controlled experiments and a/b testing.''
  \emph{Encyclopedia of machine learning and data mining}, vol.~7, no.~8, pp.
  922--929, 2017.

\bibitem{wolfert2021rate}
P.~Wolfert, J.~M. Girard, T.~Kucherenko, and T.~Belpaeme, ``To rate or not to
  rate: Investigating evaluation methods for generated co-speech gestures,''
  \emph{Proceedings of the ACM International Conference on Multimodal
  Interaction}, 2021.

\bibitem{weijters2010effect}
B.~Weijters, E.~Cabooter, and N.~Schillewaert, ``The effect of rating scale
  format on response styles: The number of response categories and response
  category labels,'' \emph{International Journal of Research in Marketing},
  vol.~27, no.~3, pp. 236--247, 2010.

\bibitem{dawes2008data}
J.~Dawes, ``Do data characteristics change according to the number of scale
  points used? an experiment using 5-point, 7-point and 10-point scales,''
  \emph{International journal of market research}, vol.~50, no.~1, pp. 61--104,
  2008.

\bibitem{jordan2020introduction}
P.~W. Jordan, \emph{An introduction to usability}.\hskip 1em plus 0.5em minus
  0.4em\relax CRC Press, 2020.

\bibitem{nakano2010estimating}
Y.~I. Nakano and R.~Ishii, ``Estimating user's engagement from eye-gaze
  behaviors in human-agent conversations,'' in \emph{Proceedings of the 15th
  international conference on Intelligent user interfaces}, 2010, pp. 139--148.

\bibitem{lemaignan2016real}
S.~Lemaignan, F.~Garcia, A.~Jacq, and P.~Dillenbourg, ``From real-time
  attention assessment to “with-me-ness” in human-robot interaction,'' in
  \emph{2016 11th ACM/IEEE International Conference on Human-Robot Interaction
  (HRI)}.\hskip 1em plus 0.5em minus 0.4em\relax Ieee, 2016, pp. 157--164.

\bibitem{hoffman2020primer}
G.~Hoffman and X.~Zhao, ``A primer for conducting experiments in human--robot
  interaction,'' \emph{ACM Transactions on Human-Robot Interaction (THRI)},
  vol.~10, no.~1, pp. 1--31, 2020.

\bibitem{liu2021speech}
Y.~Liu, G.~Mohammadi, Y.~Song, and W.~Johal, ``Speech-based gesture generation
  for robots and embodied agents: A scoping review,'' in \emph{Proceedings of
  the 9th International Conference on Human-Agent Interaction}, 2021, pp.
  31--38.

\end{thebibliography}

\end{document}